\documentclass[twocolumn,aps,pra,amsmath,amssymb,notitlepage]{revtex4-1}
\usepackage[latin9]{inputenc}
\usepackage{mathrsfs}
\usepackage{amsmath}
\usepackage{amssymb}
\usepackage{esint}

\makeatletter

\newcommand*\LyXThinSpace{\,\hspace{0pt}}

\usepackage{epsfig}
\usepackage{bm}
\usepackage{dcolumn}
\usepackage{color}

\makeatother

\begin{document}
\title{Microscopic derivation of the extended Gross-Pitaevskii equation for
quantum droplets in binary Bose mixtures}
\author{Hui Hu and Xia-Ji Liu}
\affiliation{Centre for Quantum Technology Theory, Swinburne University of Technology,
Melbourne, Victoria 3122, Australia}
\date{\today}
\begin{abstract}
An ultradilute quantum droplet is a self-bound liquid-like state recently
observed in ultracold Bose-Einstein condensates. In most previous
theoretical studies, it is described by a phenomenological low-energy
effective theory, termed as the extended Gross\textendash Pitaevskii
equation. Here, we microscopically derive the Gross\textendash Pitaevskii
equation for the condensate and also for a pairing field in an inhomogeneous
quantum droplet realized by Bose-Bose mixtures with attractive inter-species
interaction. We show that the inclusion of the pairing field is essential,
in order to have a consistent description of the droplet state. We
clarify that, the extended Gross\textendash Pitaevskii equation used
earlier should be understood as the equation of motion for the pairing
field, rather than the condensate. The fluctuations of the pairing
field give rise to low-energy collective excitations of the droplet.
We also present the Bogoliubov equations for gapless phonon modes
and gapped modes due to pairing in real space, which characterizes
single-particle-like excitations of the droplet. The equations of
motion derived in this work for the condensate and the pairing field
serve an ideal starting point to understand the structure and collective
excitations of non-uniform ultradilute quantum droplets in on-going
cold-atom experiments.
\end{abstract}
\maketitle

\section{Introduction}

One of the recent breakthroughs in ultracold atomic physics is the
realization of a self-bound liquid-like droplet state \citep{FerrierBarbut2016,Schmitt2016,Chomaz2016,Cabrera2018,Cheiney2018,Semeghini2018,Ferioli2019,Bottcher2019,DErrico2019}.
In three spatial dimensions, this new quantum state of matter arises
from a delicate balance \citep{Petrov2015} between the attractive
mean-field potential $\partial E_{\textrm{MF}}/\partial n$ and the
repulsive potential $\partial E_{\textrm{LHY}}/\partial n$ provided
by the celebrated Lee-Huang-Yang (LHY) term $E_{\textrm{LHY}}$ from
quantum fluctuations \citep{LeeHuangYang1957}, which scale like $n$
and $n^{3/2}$, respectively, as a function of the total density $n$
of the system. In almost all the theoretical studies of quantum droplets
\citep{Petrov2015,Petrov2016,Baillie2016,Wachtler2016,Baillie2017,Li2017,Cappellaro2017,Cui2018,Staudinger2018,Parisi2019,Cikojevic2019,Kartashov2019,Tylutki2020,Cikojevic2020,Wang2020,Ota2020},
the extended Gross\textendash Pitaevskii (GP) equation ($E=E_{\textrm{MF}}+E_{\textrm{LHY}}$)
\citep{Petrov2015},
\begin{equation}
i\hbar\frac{\partial}{\partial t}\psi=\left[-\frac{\hbar^{2}}{2m}\nabla^{2}+\frac{\partial E}{\partial n}\left(n=\left|\psi\right|^{2}\right)\right]\psi,\label{eq:EGPE}
\end{equation}
has been extensively used to describe the structure and dynamics of
the condensate wave $\psi$ of the droplet. This \emph{phenomenological}
low-energy effective theory is often thought to be unavoidable, since
the LHY term derived from the standard Bogoliubov theory \citep{Larsen1963}
becomes complex due to an unstable, softening phonon mode and therefore
should be amended in an empirical way \citep{Petrov2015}. As \emph{a
priori} assumption, the extended GP equation gives a useful description
of the low-lying softening phonon mode, in view of the density functional
theory.

It turns out that an amended LHY term is not the only choice. As demonstrated
in our recent work \citep{Hu2020a,Hu2020b}, the inconsistency of
a complex LHY term in the Bogoliubov theory of quantum droplets could
be alternatively removed by the inclusion of pairing between two bosons
in different species. The bosonic pairing changes the unstable softening
mode into stable \emph{gapped} mode and leads to much more accurate
ground-state energy for the droplet state, as benchmarked by the state-of-the-art
diffusion Monte Carlo simulations \citep{Parisi2019,Cikojevic2019,Cikojevic2020}.
Roughly speaking, however, the gapped mode should be considered as
single-particle excitations with relatively high energy, as we learn
from the conventional fermionic pairing theories \citep{BCS1957,Hu2006,Diener2008,Hu2008}.
Therefore, we lose the track of the low-lying collective excitations.
Immediate questions then are, how can we find those collective excitations
and how to physically interpret the phenomenological extended GP equation,
Eq. (\ref{eq:EGPE}), within the pairing theory?

In this work, we derive and discuss the equations of motion for the
condensate, the pairing field and their fluctuations in a \emph{non-uniform}
quantum droplet realized with attractive Bose-Bose mixtures, following
the framework of the previously developed microscopic pairing theory
\citep{Hu2020a,Hu2020b}. We show that the existence of the pairing
field implies additional $U(1)$ symmetry breaking for the many-body
paired bosons, i.e., the bosonic Cooper pairs. Analogous to a Bardeen\textendash Cooper-Schrieffer
(BCS) fermionic superfluid \citep{Hu2006,Diener2008}, the low-lying
collective excitations of the droplet state then should be characterized
by the pair fluctuations around the mean-field saddle-point for the
pairing field. We confirm this picture, by \emph{microscopically}
derive the extended GP equation for a large droplet, where the local
density approximation could be applied. We also present the Bogoliubov
equations for the gapless phonon mode associated with the $U(1)$
symmetry breaking of the condensate and for the gapped mode due to
the pairing. We interpret these two relatively high-lying modes as
single-particle excitations of the droplet state.

In brief, we have extended our previous work \citep{Hu2020a,Hu2020b}
to consider a finite-size quantum droplet with inhomogeneous density
distribution. Non-uniform Bogoliubov equations and extended GP equation
have been derived to describe single-particle excitations and collective
excitations of the droplet state, respectively. A self-consistent
numerical solution of these equations could provide us a more in-depth
understanding of the structure and collective oscillations of a self-bound
quantum droplet in free space.

\section{Model Hamiltonian}

We start by considering a three-dimensional homonuclear Bose-Bose
mixture such as a cloud of $^{39}$K atoms in two hyperfine states
(i.e., a $^{39}$K-$^{39}$K mixture), as in recent experiments \citep{Cabrera2018,Semeghini2018}.
For simplicity, we assume equal repulsive intra-species interactions
with strengths $g_{11}=g_{22}=g$ and attractive inter-species interactions
$g_{12}=g_{21}$, and also equal population in each species. The grand
canonical Hamiltonian of the system can then be written as,
\begin{eqnarray}
\hat{K} & = & \int d\mathbf{x}\left[\mathscr{H}_{0}+\mathscr{H}_{\textrm{intra}}+\mathscr{H}_{\textrm{inter}}\right],\\
\mathscr{H}_{0} & = & \sum_{i=1,2}\hat{\phi}_{i}^{\dagger}\left(\mathbf{x}\right)\left[-\frac{\hbar^{2}\nabla^{2}}{2m}+V_{T}\left(\mathbf{x}\right)-\mu\right]\hat{\phi}_{i}\left(\mathbf{x}\right),\\
\mathscr{H}_{\textrm{intra}} & = & \frac{g}{2}\sum_{i=1,2}\hat{\phi}_{i}^{\dagger}\left(\mathbf{x}\right)\hat{\phi}_{i}^{\dagger}\left(\mathbf{x}\right)\hat{\phi}_{i}\left(\mathbf{x}\right)\hat{\phi}_{i}\left(\mathbf{x}\right),\\
\mathscr{H}_{\textrm{inter}} & = & -\frac{\hat{\Delta}^{\dagger}\hat{\Delta}}{g_{12}}-\left[\hat{\Delta}\hat{\phi}_{1}^{\dagger}\left(\mathbf{x}\right)\hat{\phi}_{2}^{\dagger}\left(\mathbf{x}\right)+\textrm{H.c.}\right]
\end{eqnarray}
where $\hat{\phi}_{i}^{\dagger}(\mathbf{x})$ and $\hat{\phi}_{i}(\mathbf{x})$
($i=1,2$) are creation and annihilation field operators for the $i$-species
bosons with mass $m_{1}=m_{2}=m$ and with chemical potential $\mu_{1}=\mu_{2}=\mu$.
We have explicitly included an external harmonic trap $V_{T}(\mathbf{x})=m\omega^{2}\mathbf{x}^{2}/2$,
in order to account for a possible \emph{residual} weak potential
in experiments. In three dimensions, the bare interaction strengths
$g_{ij}$ are to be regularized, due to the well-known ultraviolet
divergence of the contact inter-particle interaction. We shall rewrite
them in terms of the three-dimensional $s$-wave scattering lengths
$a_{11}=a_{22}=a$ and $a_{12}$,
\begin{equation}
\frac{1}{g_{ij}}=\frac{m}{4\pi\hbar^{2}a_{ij}}-\frac{1}{\mathcal{V}}\sum_{\mathbf{k}}\frac{m}{\hbar^{2}\mathbf{k}^{2}},\label{eq:Reg3d}
\end{equation}
where $\mathcal{V}$ is the volume of the system. We note that in
$\mathscr{H}_{\textrm{inter}}$ we have taken the Hubbard-Stratonovich
transformation and have introduced a pairing field $\hat{\Delta}(\mathbf{x})$
to decouple the inter-species interaction Hamiltonian \citep{Hu2020b}.
In the weakly interacting regime, it suffices to take a \emph{static}
saddle-point solution for the pairing field. Hence, we treat $\hat{\Delta}(\mathbf{x})=\tilde{\Delta}(\mathbf{x})$
as a \emph{variational} $c$-number function. The dynamics of the
pairing field can be added back later, when we consider the fluctuations
around the saddle point.

\section{Bogoliubov theory with pairing}

We use the standard Bogoliubov theory to solve the model Hamiltonian,
in the presence of a static pairing field $\tilde{\Delta}(\mathbf{x})$.
Following Refs. \citep{Fetter1972,Griffin1996}, in the Bogoliubov
approximation we rewrite the bosonic field operators,
\begin{equation}
\hat{\phi}_{i}\left(\mathbf{x}\right)=\phi_{c}\left(\mathbf{x}\right)+\delta\hat{\phi}_{i}\left(\mathbf{x}\right),
\end{equation}
where $\delta\hat{\phi}_{i}$ is considered as a small correction
to the condensate wave-function $\phi_{c}(\mathbf{x})=e^{i\theta(\mathbf{x})}\left|\phi_{c}\left(\mathbf{x}\right)\right|$
and $\theta(\mathbf{x})$ is the phase of the condensate. For the
ground state, we take $\theta(\mathbf{x})=0$; while for a vortex
state, we set $\theta(\mathbf{x})=\varphi$ with the polar coordinate
$\varphi$. Our model Hamiltonian may then be expanded through second
order in $\delta\hat{\phi}_{i}^{\dagger}$ and $\delta\hat{\phi}_{i}$,
and the linear term vanish identically if $\phi_{c}(\mathbf{x})$
satisfies the GP equation,
\begin{equation}
\left[-\frac{\hbar^{2}\nabla^{2}}{2m}+V_{T}\left(\mathbf{x}\right)-\mu+g\left|\phi_{c}\right|^{2}\right]\phi_{c}-\tilde{\Delta}\left(\mathbf{x}\right)\phi_{c}^{*}=0.\label{eq:GPE}
\end{equation}
By introducing the notations $C(\mathbf{x})=g\left|\phi_{c}\right|^{2}$
and 
\begin{equation}
\hat{T}\left(\mathbf{x}\right)\equiv-\frac{\hbar^{2}\left[\nabla+i\nabla\theta\mathbf{\left(x\right)}\right]^{2}}{2m}+V_{T}\mathbf{\left(x\right)}-\mu,
\end{equation}
we may rewrite the GP equation into the relation for $\Delta(\mathbf{x})\equiv e^{-2i\theta}\tilde{\Delta}\left(\mathbf{x}\right)$,
\begin{equation}
\Delta\left(\mathbf{x}\right)=C\left(\mathbf{x}\right)+\left[C\left(\mathbf{x}\right)\right]^{-1/2}\hat{T}\left(\mathbf{x}\right)\left[C\left(\mathbf{x}\right)\right]^{1/2}.\label{eq:GPE2}
\end{equation}
We then obtain the truncated Bogoliubov Hamiltonian,
\begin{eqnarray}
\hat{K}_{B} & = & \sum_{i=1,2}\int d\mathbf{x}\left[\delta\hat{\phi}_{i}^{\dagger}\mathscr{L}_{0}\delta\hat{\phi}_{i}+\left(\frac{C}{2}e^{2i\theta}\delta\hat{\phi}_{i}^{\dagger}\delta\hat{\phi}_{i}^{\dagger}+\textrm{H.c.}\right)\right]\nonumber \\
 &  & -\int d\mathbf{x}\left[\left(\tilde{\Delta}\delta\hat{\phi}_{1}^{\dagger}\delta\hat{\phi}_{2}^{\dagger}+\textrm{H.c.}\right)+\frac{C^{2}}{g}+\frac{\left|\Delta\right|^{2}}{g_{12}}\right],\label{eq:BogHami}
\end{eqnarray}
where $\mathscr{L}_{0}\equiv-\hbar^{2}\nabla^{2}/(2m)+V_{T}(\mathbf{x})-\mu+2C(\mathbf{x})$.
The Bogoliubov Hamiltonian consists of a $c$-number condensate part
and a quadratic form in in $\delta\hat{\phi}_{i}$ and $\delta\hat{\phi}_{i}^{\dagger}$.
This quadratic form could be diagonalized with the linear Bogoliubov
transformation,
\begin{eqnarray}
\delta\hat{\phi}_{i}\left(\mathbf{x}\right) & = & e^{+i\theta\left(\mathbf{x}\right)}\sum_{n}\left[u_{ni}\left(\mathbf{x}\right)\hat{\alpha}_{n}+v_{ni}^{*}\left(\mathbf{x}\right)\hat{\alpha}_{n}^{\dagger}\right],\\
\delta\hat{\phi}_{i}^{\dagger}\left(\mathbf{x}\right) & = & e^{-i\theta\left(\mathbf{x}\right)}\sum_{n}\left[u_{ni}^{*}\left(\mathbf{x}\right)\hat{\alpha}_{n}^{\dagger}+v_{ni}\left(\mathbf{x}\right)\hat{\alpha}_{n}\right],
\end{eqnarray}
where $\hat{\alpha}_{n}^{\dagger}$ and $\hat{\alpha}_{n}$ are creation
and annihilation field operators of Bogoliubov quasiparticles satisfying
the usual Bose commutation relations. We can show that the truncated
Bogoliubov Hamiltonian reduces to 
\begin{eqnarray}
\hat{K}_{B} & = & -\int d\mathbf{x}\left[\frac{C^{2}\left(\mathbf{x}\right)}{g}+\frac{\left|\Delta\left(\mathbf{x}\right)\right|^{2}}{g_{12}}\right]\nonumber \\
 &  & -\int d\mathbf{x}\sum_{ni}E_{n}\left|v_{ni}\left(\mathbf{x}\right)\right|^{2}+\sum_{n}E_{n}\hat{\alpha}_{n}^{\dagger}\hat{\alpha}_{n},\label{eq:BogHamiDiag}
\end{eqnarray}
provided that the quasiparticle wave-functions $u_{ni}(\mathbf{x})$
and $v_{ni}(\mathbf{x})$ obey the coupled Bogoliubov eigenvalue equations
($i=1,2$),
\begin{eqnarray}
\mathscr{L}u_{ni}+C\left(\mathbf{x}\right)v_{ni}-\Delta\left(\mathbf{x}\right)v_{n,3-i} & = & +E_{n}u_{ni},\label{eq:BogoliubovEQ1}\\
\mathscr{L}^{*}v_{ni}+C\left(\mathbf{x}\right)u_{ni}-\Delta^{*}\left(\mathbf{x}\right)u_{n,3-i} & = & -E_{n}v_{ni},\label{eq:BogoliubovEQ2}
\end{eqnarray}
where $\mathscr{L}\equiv\hat{T}(\mathbf{x})+2C(\mathbf{x})$ is a
Hermitian operator satisfying $\int d\mathbf{x}u^{*}\mathscr{L}v=\int d\mathbf{x}v\mathscr{L}^{*}u^{*}$
\citep{Fetter1972}. It is easy to check that the wave-functions satisfy
the normalization and orthogonality conditions,
\begin{eqnarray}
\sum_{i=1,2}\int d\mathbf{x}\left[u_{ni}^{*}u_{mi}-v_{ni}^{*}v_{mi}\right] & = & \delta_{nm},\\
\sum_{i=1,2}\int d\mathbf{x}\left[u_{ni}v_{mi}-v_{ni}u_{mi}\right] & = & 0.
\end{eqnarray}
The Bogoliubov equations (\ref{eq:BogoliubovEQ1}) and (\ref{eq:BogoliubovEQ2})
have a well-known particle-hole symmetry: if $u_{ni}$ and $v_{ni}$
are a solution with energy $E_{n}$, then there is always another
solution $v_{ni}^{*}$ and $u_{ni}^{*}$ with energy $-E_{n}$. In
the diagonalized Bogoliubov Hamiltonian Eq. (\ref{eq:BogHamiDiag}),
therefore, we choose the positive eigenvalues $E_{n}\geq0$. As a
result, the thermodynamic potential at \emph{zero} temperature takes
the form,
\begin{equation}
\varOmega=-\int d\mathbf{x}\left[\frac{C^{2}}{g}+\frac{\left|\Delta\right|^{2}}{g_{12}}+\sum_{ni}E_{n}\left|v_{ni}\left(\mathbf{x}\right)\right|^{2}\right],
\end{equation}
from which, we may determine the variational pairing field $\Delta(\mathbf{x})$
through the \emph{functional} minimization, i.e., 
\begin{equation}
\frac{\delta\varOmega\left[\mu,\Delta\left(\mathbf{x}\right)\right]}{\delta\Delta(\mathbf{x})}=0.
\end{equation}
It is worth noting that the LHY contribution $\varOmega_{\textrm{LHY}}=-\int d\mathbf{x}\sum_{ni}E_{n}\left|v_{ni}(\mathbf{x})\right|^{2}$
is formally divergent. This divergence can be removed by regularizing
the bare interaction strengths $g$ and $g_{12}$. In practice, we
introduce a high-energy cut-off energy $E_{c}$, above which the discreteness
of the energy spectrum $E_{n}$ is no longer important and we semi-classically
solve the Bogoliubov equations under the local density approximation
to obtain $u_{\mathbf{k}i}(\mathbf{x})$, $v_{\mathbf{k}i}(\mathbf{x})$
and $E_{\mathbf{k}}$ \citep{Liu2007}. We then rewrite the thermodynamic
potential into two parts, $\varOmega=\int d\mathbf{x}[\varOmega_{d}(\mathbf{x})+\varOmega_{c}(\mathbf{x})]$,
where 
\begin{eqnarray}
\varOmega_{d} & = & -\frac{m}{4\pi\hbar^{2}}\left(\frac{C^{2}}{a}+\frac{\left|\Delta\right|^{2}}{a_{12}}\right)-\sum_{i,E_{n}<E_{c}}E_{n}\left|v_{ni}\right|^{2},\label{eq:OmegaDiscrete}\\
\varOmega_{c} & = & \sum_{\mathbf{k}}\frac{m\left(C^{2}+\left|\Delta\right|^{2}\right)}{\hbar^{2}\mathbf{k}^{2}}-\sum_{i,E_{\mathbf{k}}\geq E_{c}}E_{\mathbf{k}}\left|v_{\mathbf{k}i}\right|^{2}.\label{eq:OmegaContinuous}
\end{eqnarray}
For a given chemical potential $\mu$, the GP equation (\ref{eq:GPE}),
the Bogoliubov equations (\ref{eq:BogoliubovEQ1}) and (\ref{eq:BogoliubovEQ2}),
and the thermodynamic potential Eqs. (\ref{eq:OmegaDiscrete}) and
(\ref{eq:OmegaContinuous}) form a \emph{closed} set of equations
to determine $C(\mathbf{x})$ and $\Delta(\mathbf{x})$, and consequently
the condensate wave-function $\left|\phi_{c}(\mathbf{x})\right|\propto\sqrt{C(\mathbf{x})}$
and the Bogoliubov spectrum $E_{n}$. This is the first key result
of our work. The total number of atoms can be calculated from the
thermodynamic potential relation $N=-\partial\varOmega/\partial\mu$,
which provides the normalization to $\phi_{c}(\mathbf{x})$. We note
that, if we neglect the quantum depletion, which is small in the weakly
interacting regime, we may write down directly $\left|\phi_{c}(\mathbf{x})\right|=\sqrt{mC(\mathbf{x})/(4\pi\hbar^{2}a)}$.

For simplicity, from now on we focus on the ground state with a phase
$\theta(\mathbf{x})=0$. Our discussion given below can be easily
extended to the general case with a nonzero phase factor $\theta(\mathbf{x})\neq0$
and we will put $\theta(\mathbf{x})$ back when it is needed.

\section{Bulk properties of quantum droplets}

For a large droplet in the absence of any external potential and in
the ground state, the function $C(\mathbf{x})$ and the pairing field
$\Delta(\mathbf{x})$ are essentially \emph{real} constant, except
at the edge of the droplet. Thus, if we neglect the edge effect, the
GP equation (\ref{eq:GPE}) gives the relation $C=\mu+\Delta$. The
Bogoliubov equations (\ref{eq:BogoliubovEQ1}) and (\ref{eq:BogoliubovEQ2})
in momentum space take the form,
\begin{equation}
\left[\begin{array}{cccc}
B_{\mathbf{k}} & 0 & C & -\Delta\\
0 & B_{\mathbf{k}} & -\Delta & C\\
C & -\Delta & B_{\mathbf{k}} & 0\\
-\Delta & C & 0 & B_{\mathbf{k}}
\end{array}\right]\left[\begin{array}{c}
u_{\mathbf{k}1}\\
u_{\mathbf{k}2}\\
v_{\mathbf{k}1}\\
v_{\mathbf{k}2}
\end{array}\right]=E_{\mathbf{k}}\left[\begin{array}{c}
+u_{\mathbf{k}1}\\
+u_{\mathbf{k}2}\\
-v_{\mathbf{k}1}\\
-v_{\mathbf{k}2}
\end{array}\right],
\end{equation}
where $B_{\mathbf{k}}\equiv\varepsilon_{\mathbf{k}}+C+\Delta$ with
$\varepsilon_{\mathbf{k}}=\hbar^{2}\mathbf{k}^{2}/(2m)$. We then
obtain two Bogoliubov spectra,

\begin{eqnarray}
E_{-}(\mathbf{k}) & = & \sqrt{\varepsilon_{\mathbf{k}}\left(\varepsilon_{\mathbf{k}}+2C+2\Delta\right)},\label{eq: Emk}\\
E_{+}(\mathbf{k}) & = & \sqrt{\left(\varepsilon_{\mathbf{k}}+2C\right)\left(\varepsilon_{\mathbf{k}}+2\Delta\right)}.\label{eq:Epk}
\end{eqnarray}
For both dispersions, the wave-functions $u_{\mathbf{k}i}(\mathbf{x})=u_{\mathbf{k}i}e^{i\mathbf{k}\cdot\mathbf{x}}/\sqrt{\mathcal{V}}$
and $v_{\mathbf{k}i}(\mathbf{x})=v_{\mathbf{k}i}e^{i\mathbf{k}\cdot\mathbf{x}}/\sqrt{\mathcal{V}}$
are given by,
\begin{align}
u_{\mathbf{k}1}^{2} & =u_{\mathbf{k}2}^{2}=\frac{1}{4}\left(\frac{B_{\mathbf{k}}}{E_{\mathbf{k}}}+1\right),\label{eq:uk2}\\
v_{\mathbf{k}1}^{2} & =v_{\mathbf{k}2}^{2}=\frac{1}{4}\left(\frac{B_{\mathbf{k}}}{E_{\mathbf{k}}}-1\right),\label{eq:vk2}
\end{align}
from which, we determine the thermodynamic potential,

\begin{eqnarray}
\frac{\varOmega}{\mathcal{V}} & = & -\frac{m}{4\pi\hbar^{2}}\left[\frac{C^{2}}{a}+\frac{\Delta^{2}}{a_{12}}\right]+\frac{1}{2}\sum_{\mathbf{k}}\left[E_{+}\left(\mathbf{k}\right)+E_{-}\left(\mathbf{k}\right)\right.\nonumber \\
 &  & \left.-2\left(\varepsilon_{\mathbf{k}}+C+\Delta\right)+\frac{C^{2}+\Delta^{2}}{\varepsilon_{\mathbf{k}}}\right].
\end{eqnarray}
This result was obtained in our previous works using a path-integral
functional approach \citep{Hu2020a,Hu2020b}. By integrating over
the momentum $\mathbf{k}$, we arrive at the expression,
\begin{equation}
\frac{\varOmega}{\mathcal{V}}=-\frac{m}{4\pi\hbar^{2}}\left[\frac{C^{2}}{a}+\frac{\Delta^{2}}{a_{12}}\right]+\frac{8m^{3/2}C^{5/2}}{15\pi^{2}\hbar^{3}}\mathcal{G}_{3}\left(\frac{\Delta}{C}\right),\label{eq:OmegaUniform}
\end{equation}
where $\mathcal{G}_{3}(\alpha)\equiv(1+\alpha)^{5/2}+h_{3}(\alpha)$
with $h_{3}(\alpha)\equiv(15/4)\int_{0}^{\infty}dt\sqrt{t}[\sqrt{(t+1)(t+\alpha)}-(t+1/2+\alpha/2)+(1-\alpha)^{2}/(8t)]$
. Near the equilibrium density of quantum droplets, the chemical potential
$\left|\mu\right|$ is typically much smaller than $C$ and $\Delta$
\citep{Hu2020a,Hu2020b}. We may then expand $\varOmega\left(\mu\right)$
in powers of $\mu$, $\varOmega\left(\mu\right)=\varOmega^{(0)}+\mu\varOmega^{(1)}+\cdots$,
where 
\begin{eqnarray}
\frac{\varOmega^{(0)}}{\mathcal{V}} & = & -\frac{m}{4\pi\hbar^{2}}\left(\frac{1}{a}+\frac{1}{a_{12}}\right)\Delta^{2}+\frac{32\sqrt{2}m^{3/2}}{15\pi^{2}\hbar^{3}}\Delta^{5/2},\\
\frac{\varOmega^{(1)}}{\mathcal{V}} & = & -\frac{m}{2\pi\hbar^{2}a}\Delta+\frac{8\sqrt{2}m^{3/2}}{3\pi^{2}\hbar^{3}}\Delta^{3/2}.
\end{eqnarray}
By taking the derivative $-\partial\varOmega/\partial\mu$ \citep{NoteCriticalMu},
we obtain the pairing gap 
\begin{equation}
\Delta\simeq\frac{2\pi\hbar^{2}a}{m}n\left[1+\eta\right],\label{eq:DeltaN}
\end{equation}
 and the total energy per unit volume $E/\mathcal{V}=\mu n+\varOmega/\mathcal{V}$,
\begin{eqnarray}
\frac{E}{\mathcal{V}} & \simeq & -\frac{\pi\hbar^{2}}{m}\left(a+\frac{a^{2}}{a_{12}}\right)n^{2}\left(1+2\eta\right)\nonumber \\
 &  & +\frac{256\sqrt{\pi}}{15}\frac{\hbar^{2}a^{5/2}}{m}n^{5/2}\left(1+\frac{5}{2}\eta\right),\label{eq:EnergyN}
\end{eqnarray}
where the correction factor $\eta\equiv32\sqrt{na^{3}}/(3\sqrt{\pi})\ll1$
comes from the second term in $\varOmega^{(1)}/\mathcal{V}$. This
small correction is absent if we approximate $C\simeq\Delta$ in the
LHY thermodynamic potential (i.e., the second term on the right-hand-side
of Eq. (\ref{eq:OmegaUniform})).

\section{Large droplets within the local density approximation}

To take into account the edge effect for a \emph{large} quantum droplet,
we may take the local density approximation, by assuming very slowly
varying $C(\mathbf{x})$ and $\Delta(\mathbf{x})$ in real space.
That is, $C(\mathbf{x})$ and $\Delta(\mathbf{x})$ change very slowly
at the scale of the healing length $\xi_{c}=(8\pi na)^{-1/2}$, which
is set by the condition $\hbar^{2}/(2m\xi_{c}^{2})\sim4\pi\hbar^{2}an/m$
\citep{Dalfovo1999}. This amounts to neglecting all the discrete
energy levels in the summation of Eq. (\ref{eq:OmegaDiscrete}). In
other words, we treat all the wave-functions classically as plane-waves
in a locally uniform cell (located at the position $\mathbf{x}$)
with a well-defined wave-vector $\mathbf{k}$. The quasi-particle
amplitudes $u_{\mathbf{k}i}$ and $v_{\mathbf{k}i}$ and the corresponding
energy level $E_{\mathbf{k}}$ (including the two branches $E_{-}(\mathbf{k})$
and $E_{+}(\mathbf{k})$) can then be obtained from Eqs. (\ref{eq:uk2}),
(\ref{eq:vk2}), (\ref{eq: Emk}) and (\ref{eq:Epk}), respectively,
with the position dependence is explicitly kept by using the slowly
spatially varying function $C(\mathbf{x})$ and the pairing field
$\Delta(\mathbf{x})$. The neglect of the discrete energy levels is
equivalent to setting the cut-off energy $E_{c}=0$ in Eqs. (\ref{eq:OmegaDiscrete})
and (\ref{eq:OmegaContinuous}), and consequently at each position
$\mathbf{x}$ we have,
\begin{equation}
\varOmega_{d}=-\frac{m}{4\pi\hbar^{2}}\left[\frac{C^{2}\left(\mathbf{x}\right)}{a}+\frac{\Delta^{2}\left(\mathbf{x}\right)}{a_{12}}\right],\label{eq:OmegaDiscrete2}
\end{equation}
and
\begin{equation}
\varOmega_{c}=\sum_{\mathbf{k}}\frac{C^{2}\left(\mathbf{x}\right)+\Delta^{2}\left(\mathbf{x}\right)}{2\varepsilon_{\mathbf{k}}}-\sum_{E_{\mathbf{k}}}\frac{B_{\mathbf{k}}\left(\mathbf{x}\right)-E_{\mathbf{k}}\left(\mathbf{x}\right)}{2},\label{eq:OmegaContinuous2}
\end{equation}
where we have treated $\Delta(\mathbf{x})$ as a \emph{real} and non-negative
function, since it only differs slightly from $C(\mathbf{x})$. We
note that, $B_{\mathbf{k}}\left(\mathbf{x}\right)$ and $E_{\mathbf{k}}(\mathbf{x})$
in the expression of $\varOmega_{c}$ should be understood as the
local dispersion relations at the position $\mathbf{x}$ with $C(\mathbf{x})$
and $\Delta(\mathbf{x})$, and we also need to consider both dispersions
$E_{-}(\mathbf{k};\mathbf{x})$ and $E_{+}(\mathbf{k};\mathbf{x})$
. The summation in $\varOmega_{c}$ over the momentum $\mathbf{k}$
at the position $\mathbf{x}$ is easy to carry out. Following the
derivation in Eq. (\ref{eq:OmegaUniform}), we obtain
\begin{equation}
\varOmega_{c}=\frac{8m^{3/2}}{15\pi^{2}\hbar^{3}}\left[C\left(\mathbf{x}\right)+\Delta\left(\mathbf{x}\right)\right]{}^{5/2},
\end{equation}
where we have used the fact that $C(\mathbf{x})\simeq\Delta(\mathbf{x})$,
so the function $h_{3}(C/\Delta)\propto[C(\mathbf{x})-\Delta(\mathbf{x})]^{2}/\Delta^{2}(\mathbf{x})\ll1$
can be safely neglected \citep{Hu2020a}. By adding the contribution
$\varOmega_{d}$ in Eq. (\ref{eq:OmegaDiscrete2}), we find that
\begin{eqnarray}
\varOmega & = & -\int d\mathbf{x}\frac{m}{4\pi\hbar^{2}}\left[\frac{C^{2}\left(\mathbf{x}\right)}{a}+\frac{\Delta^{2}\left(\mathbf{x}\right)}{a_{12}}\right]\nonumber \\
 &  & +\int d\mathbf{x}\frac{8m^{3/2}}{15\pi^{2}\hbar^{3}}\left[C\left(\mathbf{x}\right)+\Delta\left(\mathbf{x}\right)\right]{}^{5/2}.
\end{eqnarray}

By taking the functional derivative $\delta\varOmega[\Delta(\mathbf{x})]/\delta\Delta=0$,
we find that,
\begin{equation}
\left[\frac{C}{a}\frac{\delta C}{\delta\Delta}+\frac{\Delta}{a_{12}}\right]-\frac{16\sqrt{m}}{3\pi\hbar}\left[C+\Delta\right]^{3/2}\left(\frac{\delta C}{\delta\Delta}+1\right)=0.\label{eq:DOmegaDdelta}
\end{equation}
From the GP equation (\ref{eq:GPE2}), the function $C(\mathbf{x})$
is related to the pairing field $\Delta(\mathbf{x})$,
\begin{equation}
C\left(\mathbf{x}\right)\simeq\Delta\left(\mathbf{x}\right)-\left[\Delta\left(\mathbf{x}\right)\right]^{-1/2}\hat{T}\left(\mathbf{x}\right)\left[\Delta\left(\mathbf{x}\right)\right]^{1/2}.
\end{equation}
As $\Delta(\mathbf{x})$ is a very slowly varying function for a large
droplet, to a good approximation we may take $\delta C/\delta\Delta=1$
and also set $C(\mathbf{x})\simeq\Delta(\mathbf{x})$ in the second
term of Eq. (\ref{eq:DOmegaDdelta}). The latter is equivalent to
neglecting the small correction factor $\eta$ in Eq. (\ref{eq:DeltaN})
and Eq. (\ref{eq:EnergyN}), as we discussed earlier. By further inspired
by the relation (\ref{eq:DeltaN}) to introduce $\Phi^{2}(\mathbf{x})\equiv[m/(2\pi\hbar^{2}a)]e^{2i\theta(\mathbf{x})}\Delta(\mathbf{x})=[m/(2\pi\hbar^{2}a)]\tilde{\Delta}(\mathbf{x})$,
we rewrite Eq. (\ref{eq:DOmegaDdelta}) into the form,
\begin{equation}
\left[\hat{T}-\frac{2\pi\hbar^{2}}{m}\left(a+\frac{a^{2}}{a_{12}}\right)\left|\Phi\right|^{2}+\frac{128\sqrt{\pi}}{3}\frac{\hbar^{2}a^{5/2}}{m}\left|\Phi\right|^{3}\right]\Phi=0.\label{eq:EGPEOurs}
\end{equation}
By recalling $\hat{T}=-\hbar^{2}\nabla^{2}/(2m)-\mu=-\hbar^{2}\nabla^{2}/(2m)-i\hbar\partial_{t}$
in the absence of external potential, the above equation is precisely
the extended GP equation (\ref{eq:EGPE}), once we take Eq. (\ref{eq:EnergyN})
with $\eta=0$ as the density functional $E(n)$. Hence, we have microscopically
derived the extended GP equation, under the condition (i.e., within
the local density approximation) that it is applicable.

To check the self-consistency of the local density approximation,
it is useful to note that, from Eq. (\ref{eq:EGPEOurs}) the droplet
typically changes at the length scale \citep{Petrov2015} 
\begin{equation}
\xi=\frac{64\sqrt{6}}{5\pi}\left(1+\frac{a}{a_{12}}\right)^{-3/2}a
\end{equation}
with an equilibrium density 
\begin{equation}
n=\frac{25\pi}{16384}\left(1+\frac{a}{a_{12}}\right)^{2}a^{-3}.
\end{equation}
Therefore, we find the ratio,
\begin{equation}
\frac{\xi}{\xi_{c}}=2\sqrt{3}\left(1+\frac{a}{a_{12}}\right)^{-1/2},
\end{equation}
which is about $11.5$ under the typical experimental condition $a_{12}\simeq-1.1a$
\citep{Cabrera2018,Semeghini2018}. As a result of $\xi\gg\xi_{c}$,
the local density approximation is well satisfied for the experimentally
realized Bose droplets \citep{Cabrera2018,Semeghini2018}.

Our derivation clearly shows that the wave-function in the extended
GP equation (\ref{eq:EGPE}) represents the pairing field $\tilde{\Delta}(\mathbf{x})$,
rather than the condensate wave-function $\phi_{c}(\mathbf{x})$,
as one may naïvely anticipate. The latter satisfies instead the ordinary
GP equation, as given in Eq. (\ref{eq:GPE}). This clarification is
another key result of our work.

In our microscopic pairing theory, the quantum droplet could be viewed
as a mixture of bosonic atoms and loosely-bound many-body bosonic
Cooper pairs, analogous to a two-component Fermi superfluid where
fermions and loosely-bound fermionic Cooper pairs co-exist \citep{BCS1957,Hu2006,Diener2008}.
Eq. (\ref{eq:EGPEOurs}) derived here can therefore be regarded as
the bosonic counterpart of the well-known Ginzburg\textendash Landau
equation for the BCS pairing order parameter \citep{SadeMelo1993}.
Physically, there are two $U(1)$ symmetry breakings, one is associated
with the condensate wave-function $\phi_{c}(\mathbf{x})$ and another
with the pairing field $\tilde{\Delta}(\mathbf{x})$. As in a fermionic
superfluid, low-energy \emph{collective} excitations correspond to
the fluctuations around the static saddle-point solution $\tilde{\Delta}(\mathbf{x})$
and can be studied by linearizing the extended GP equations (\ref{eq:EGPE})
or (\ref{eq:EGPEOurs}) with 
\begin{equation}
\Phi\left(t\right)=e^{-i\mu t/\hbar}\left[\Phi_{0}+\sum_{n}\left(U_{n}e^{-i\omega_{n}t}+V_{n}^{*}e^{+i\omega_{n}t}\right)\right],
\end{equation}
to the first order in fluctuations $U_{n}(\mathbf{x})$ and $V_{n}(\mathbf{x})$.
This leads to the Bogoliubov equations for pairing fluctuations. In
contrast, the quasi-particles described by the Bogoliubov equations
in (\ref{eq:BogoliubovEQ1}) and (\ref{eq:BogoliubovEQ2}) should
be understood as \emph{single-particle} excitations of bosonic atoms.

It is worth noting that, in a scalar weakly-interacting Bose gas collective
excitations and single-particle excitations are strongly correlated,
due to the existence of the condensate, so the Bogoliubov quasi-particles
are generally treated as collective excitations \citep{Griffin1993,Liu2004}.
The situation in quantum droplets is less clear, as a result of the
bosonic pairing. A careful identification of collective excitations
and single-particle excitations is therefore needed, by calculating
of density-density correlation functions, presumably within the random-phase-approximation
\citep{Liu2004}.

\section{Conclusions}

In summary, we have constructed a microscopic pairing theory for describing
the non-uniform states of quantum droplets realized with an attractive
Bose-Bose mixture in three dimensions. We have pointed out that the
droplet can be viewed as a coherent mixture of bosonic atoms and loosely-bound
bosonic Cooper pairs, both of which are Bose-condensed. We have presented
the equations of motion for the atomic condensate and the pairing
field of Cooper pairs, as well as the fluctuations around them. The
resulting closed set of equations (\ref{eq:GPE}), (\ref{eq:BogoliubovEQ1}),
(\ref{eq:BogoliubovEQ2}), (\ref{eq:OmegaDiscrete}) and (\ref{eq:OmegaContinuous})
forms the basis to investigate the structure and dynamics of quantum
droplets in future studies.

By using these equations for a large quantum droplet, where the spatial
variation in the condensate wave-function and the pairing field is
small, we have microscopically derived the extended Gross-Pitaevskii
equation, which has been previously used a phenomenological low-energy
effective theory. We have clarified that the extended Gross-Pitaevskii
equation describes the pairing field, instead of the condensate wave-function
as one may naïvely expect. This clarification would be important for
properly distinguishing the collective excitations and single-particle
behaviors in the intriguing new quantum state of self-bound liquid-like
droplets.

We note finally that, in practice the functional minimization of the
thermodynamic potential with respect to the variational pairing field
is difficult to carry out numerically. For the most relevant case
of a self-bound \emph{spherical} quantum droplet in free space, therefore
it is useful to parameterize the pairing field $\Delta(r)$ by using
a few variational parameters. For example, for a large droplet we
may consider a flat-top distributed droplet with the form
\begin{equation}
\Delta(r)=\Delta_{0}\left[1+\frac{\cosh\left(\kappa r\right)}{\cosh\left(\kappa R\right)}\right]^{-1},\label{eq:PairingFieldVariational}
\end{equation}
where the parameter $\kappa$ controls the shape of the droplet at
the edge $r\sim R$. The initial input for the variational parameters
$\Delta_{0}$, $\kappa$ and $R$ can be obtained by solving the extended
Gross-Pitaevskii Eq. (\ref{eq:EGPEOurs}). For the given pairing field
$\Delta(r)$ in Eq. (\ref{eq:PairingFieldVariational}) and chemical
potential $\mu<0$ (for the droplet state), we first solve the nonlinear
Gross-Pitaevskii Eq. (\ref{eq:GPE}) for $C(r)$ by using the standard
iterative algorithm \citep{Pu1998}. As the effective potential in
the Gross-Pitaevskii equation, i.e., $V_{\textrm{eff}}=-\mu-\Delta(r)$
always curves upwards, we could numerically find a stable solution
for the function $C(r)$, which acquires a flat-top distribution with
$C(r=0)=\mu+\Delta_{0}$ and vanishes at $r\sim R$. In the second
step, we solve the coupled Bogoliubov equations, Eqs. (\ref{eq:BogoliubovEQ1})
and (\ref{eq:BogoliubovEQ2}), for the quasi-particle energies and
wave-functions, following the work by Hutchinson, Zaremba and Griffin
\citep{Hutchinson1997}. We then calculate the total thermodynamic
potential $\varOmega$ by using Eqs. (\ref{eq:OmegaDiscrete}) and
(\ref{eq:OmegaContinuous}). In this way, we obtain $\varOmega$ as
functions of $\Delta_{0}$, $\kappa$ and $R$. The minimization then
leads to an optimum set of the variational parameters, improving the
initial guess from the solution of Eq. (\ref{eq:EGPEOurs}).
\begin{acknowledgments}
This research was supported by the Australian Research Council's (ARC)
Discovery Program, Grant No. DP170104008 (H.H.) and Grant No. DP180102018
(X.-J.L).
\end{acknowledgments}


\begin{thebibliography}{10}
\bibitem{NoteCriticalMu}We note that, if we minimize $\varOmega(\mu,\Delta)$
with respect to $\Delta$, we obtain the condition, $\mu\simeq(1+a/a_{12})\Delta+[32\sqrt{2}m^{1/2}a/(3\pi\hbar)]\Delta^{3/2}$,
which is bound from below. The low bound $\mu_{c}=-(\pi^{2}/768)(1+a/a_{12})^{3}[\hbar^{2}/(2ma^{2})]$
is the critical chemical potential, below which one can no longer
find the local minimum in $\varOmega(\mu,\Delta)$. Therefore, $\mu_{c}$
is the critical chemical potential for the droplet-to-gas transition,
which happens at sufficiently low density.
\end{thebibliography}
\end{document}